
\input jnl.tex
\rightline {hep-th/9307059}
\rightline {NSF-ITP-93-57}
\def\d{^\dagger }
\title BLACK HOLE STATISTICS

\author Andrew Strominger
\affil Institute for Theoretical Physics
  and
 Department of Physics,
 University of California
Santa Barbara, CA 93106

\abstract{
The quantum statistics of charged, extremal black holes is
investigated beginning with the hypothesis that the quantum state is a
functional on the space of closed three-geometries, with each black hole
connected to an oppositely charged black hole
through a spatial wormhole. From this starting point a
simple argument is given that a collection of extremal black holes obeys
neither
Bose nor Fermi statistics. Rather they obey an exotic variety of
particle statistics
known as ``infinite statistics'' which resembles that of
distinguishable particles and is realized
by a $q$-deformation of the quantum commutation relations.}

\vfill\eject
\doublespace
Semiclassical calculations [\cite{hwk}]
indicate that  the mass of a charge $Q$ black hole
will decrease via Hawking radiation until it reaches a critical value
proportional to $Q$. The resulting ``extremal" black hole appears to be a
quantum mechanically stable object\footnote{$\dagger$}{In
 a world in which there
are no elementary particles with mass less than
$QM_{planck}$.}. Extremal black holes
have been found to be a useful item in the gedanken laboratory for studying the
quantum mechanics of black holes. Their utility derives from the fact that, for
large $Q$, they are macroscopic objects whose behavior should be governed by
(hopefully) well-understood laws of low-energy physics. Thus the
short-distance problems of quantum gravity might be divorced from the quantum
puzzles of black holes. In the  last several
years there has accordingly been much
progress in understanding the scattering of low-energy particles by extremal
black holes[\cite{lyr}].

In this paper we shall consider a different process: the scattering of two
extremal black holes. There are many interesting aspects of this problem, but
we shall consider here only the zeroth-order question ``Do they scatter as
bosons, fermions, or something else?". We shall argue that at least in
some cases the answer is ``something else''.

In non-gravitational field theories, it was shown long ago by Finkelstein and
Rubenstein[\cite{frb,bal}] that soliton statistics can be determined from the
fact
that the quantum state is a functional (or more generally sectional) on the
space of field configurations. For example, double exchange of a pair
of skyrmions is an
operation which is continuously deformable to
the identity, while single exchange is an operation
which is continuously deformable to a $2\pi$ rotation  of a single skyrmion.
Thus the exchange operator
$\cal E$ must have eigenvalues either plus or minus one (since it squares to
the
identity) and the solitons may correspondingly be either bosons or fermions.
Since $\cal E$ is continuously deformable to a $2\pi$ rotation of a single
skyrmion, the eigenvalue is connected with the angular momentum of the skyrmion
wave
function, in accord with the expected spin-statistics relation.

If the soliton has no internal states, then the exchange operator $\cal E$ will
commute with every operator in the theory simply because it does not change the
physical field configuration.
This  has the consequence that the eigenspaces of $\cal E$ are
superselection sectors: bosons are forever
bosons and fermions are forever fermions. If the soliton does have
internal states then $\cal E$ may or may not change the field configuration,
depending on whether or not the solitons are in the same state. $\cal E$ will
then
obey a composition law with those operators which act on the internal state
reflecting the statistics of identical particles with internal degrees of
freedom.

The application of these ideas to quantum gravity is developed in a series of
beautiful papers by Friedman and Sorkin[\cite{fsk}]. In quantum gravity, the
wave function must be invariant under all diffeomorphisms which are
asymptotically trivial and
can be continuously deformed to the identity. The square
${\cal E}^2$ of the exchange operator acting on two identical solitons is in
this
category, but  $\cal E$ itself is in general (depending on the soliton in
question)
a non-trivial ``large" $Z_2$ diffeomorphism.
These considerations alone may allow either Bose or Fermi statistics but,
because observables must commute with large diffeomorphisms, the statistics
form
superselection sectors. In some cases the statistics can be fixed by
consideration of the soliton creation process[\cite{srdw}].

We wish to apply these ideas to the problem of charged extremal black hole
statistics. The first issue is the relevant space of three-geometries.

Perhaps
the first possibility which comes to mind is to consider
three-geometries with a
boundary at each black hole horizon. One must then decide whether or not these
boundaries are distinguishable, which is completely equivalent to deciding the
statistics. So nothing is learned in this approach.

It is preferable to formulate the problem in terms
of three-geometries without boundaries. Because of flux conservation, the two
sphere surrounding the black hole cannot be topologically trivial (assuming
there are no charged sources). In this paper we will
investigate the consequences of the assumption
 that the flux reemerges elsewhere in our (or possibly another) universe out of
an oppositely
charged extremal black hole. The relevant space of three geometries is then
$R^3$ with $N$ handles whose cross sections are two-spheres, corresponding to
$N$ positively-charged and $N$ negatively-charged black holes. It is
worth noting that this formulation is consistent with semiclassical
instanton calculations [\cite{gast,bol}], which describe creation of charged
black hole
pairs connected by a wormhole in an electromagnetic field.

Yet another possibility - briefly discussed below - is to allow the flux to end
at an `t~Hooft-Polyakov
monopole or an electric charge inside the horizon. This formulation is
inherently more complex as it necessarily
involves time-dependent three  geometries.
However, because these descriptions differ only behind a horizon,
one may  hope that they
lead to the same observable consequences.

We further want to consider only extremal black holes with a unique ground
state, so that the exchange operator is a diffeomorphism. This is quite
different from (though of course relevant to) the situation considered
in [\cite{bol}], wherein extremal black holes were excited in various ways by
low-energy scattering of massless fermions. In a theory
with no massless fermions or scalars, there is an energy barrier to throwing
matter into the black hole, and it
should be possible to scatter black holes at sufficiently low energies without
exciting any internal degrees of freedom.


The basic point is now very simple. Black hole exchange is the exchange of two
wormhole ends. This  results in a new three-geometry: it is not a
diffeomorphism. Quantum mechanics does {\it not}
require that the wave function should have any
particular symmetry properties under this operation. The wave function of $N$
like-charge black holes will be a general function $\psi(x_1,\dots x_N)$
of their
$N$ postions.
Extremal black hole scattering will resemble that of distinguishable particles:
It is also the case that the wave function for $N$ identical particles with
more than $N$ internal states is a general function
$\psi(x_1,\dots x_N)$ if none of the
$N$ particles are in the same internal state. The low-energy experimentalist
who discovers the extremal black holes might believe that he has found a new
type of massive particle with a large number of degenerate internal states.
However, no
matter how long she/he searches, two particles in the same internal state
({\it i.e.} that scatter like identical particles) will
never be found.

While black hole exchange is not a large diffeomorphism, wormhole exchange is.
This involves the simultaneous exchange of a negatively-charged and a
positively-charged black hole pair. A priori the eigenvalue of the wormhole
exchange operator
${\cal E}_w$ may have either sign. However, as in previously
studied examples[\cite{srdw}], the sign is fixed by consistency with
the rules for pair creation.  In [\cite{gast}] an instanton was found
describing single wormhole creation in a magnetic field. To get the creation
rate one
must exponentiate the single instanton. In so doing one counts only once the
four-geometries which differ by wormhole exchange on the final three-geometry.
Thus it is implicit in this calculation---which led to reasonable
results---that ${\cal E}_w$ has eigenvalue plus one.  We shall henceforth
assume
the wormholes are bosons\footnote{*}{ This is also
consistent with the spin-statistics relation, since the single instanton
creates the
wormhole in a state invariant under $2\pi$ rotations.}.

One might expect that the Bose nature of black hole pairs would lead to
observable consequences.
In order to scatter two wormholes one must first locate the two charge $-Q$
partners of two charge $+Q$ black
holes. One might attempt to do this by shining
a flashlight in one end of the wormhole and seeing where the light reemerges.
Of course classically this is prohibited by causality: there is a horizon in
the middle of the wormhole\footnote{$\dagger$}{Although the spacelike slice
containing the wormhole may cross a pair of horizons without entering into a
region of trapped surfaces, as in the $t=0$ slice of maximally extended
Schwarzchild.}. This is a special case of more general
``topological censorship" theorems [\cite{tc}]. The gist of these theorems is
that classically it is impossible for an external observer to detect
non-trivial
topology.

If topological censorship were not generally
valid, many pathologies would arise. For
example wormhole traversal might be used for time travel[\cite{mty}].
Thus it is natural to
assume that {\it quantum} topological censorship is valid as well. In this case
one can
never locate the other end of a given extremal black hole and observe Bose
scattering of wormholes.

One might be concerned that if extremal
black holes are quantum mechanically similar to elementary particles with an
infinite number of internal states that they will share with the latter object
an infinite pair-production rate. Our description of extremal black holes is
consistent with---indeed was inspired by---the description of quantum gravity
transition amplitudes as sums over {\it inequivalent\/} four geometries.
Euclidean instanton methods were used in [\cite{bol,gast}] to argue that this
prescription leads to a finite (and semiclassically computable) pair-production
rate. Thus - at least in this
description - the quantum mechanics of
extremal black holes differs in this regard
from that of an object with an infinite number of internal states.
Indeed, the experimentalist who tires in her/his search for two objects
that interfere like identical particles might learn that he/she is on
the wrong track by measuring the pair-production rate.

The arguments given herein mesh well with, and are in a sense an extension of,
those given in [\cite{bol}]. In [\cite{bol}] we considered (in a theory with
massless fermions) the possibility that information lost in low-energy
particle-hole scattering might be stored in the form of an infinite degeneracy
of zero-energy  black hole states in which the matter fields  are excited
behind the horizon. One might expect that this information could be recovered
by quantum interference experiments. It was argued  [\cite{bol}] that in fact
causality dooms such experiments to failure, and black holes will
always scatter as distinguishable particles, even if they are initially in the
same zero-energy state. In this paper we have argued that
this is the case even for extremal black holes with no internal excitations,
from which this last conclusion would certainly follow.

Let us now briefly consider a description of extremal magnetic black holes in
which there are no wormholes, and magnetic flux is terminated at an
`t~Hooft-Polyakov monopole. In a theory with
$M_{{\rm GUT}}>M_{{\rm Planck}}$\footnote{$\dagger$}{Although this makes sense
classically, it is far from clear that there is any quantum mechanical meaning
to a particle with Compton wavelength less than its Schwarzchild radius!}, a
smooth, slowly varying field configuration with non-zero magnetic charge will
in general collapse to form a magnetic black hole, which will then Hawking
radiate down to extremality. The exterior of the black hole configuration will
then be
static, but the interior will continue to evolve.

In this system the topology of a spacelike slice with a collection of black
holes is simply $R^3$, provided we continue the spacelike slice inside the
horizon in such a way so as to avoid any singularities. An extremal black hole
is therefore represented as an object with many internal states, associated
with the degrees of freedom inside the horizon. In the scattering of two
like-charge extremal black holes, those internal states cannot be observed and
should therefore be traced over. In general the scattering will therefore
resemble non-identical particles. However, a better understanding of the
dynamics is required to understand to what extent this description of black
hole scattering
is consistent with the previous one.


Ultimately one might hope to obtain a quantum field theoretic description of
extremal black holes in which they are treated as point particles.
A classification of all possible particle statistics consistent with general
principles of quantum
field theory was obtained by Doplicher, Haag and Roberts[\cite{dhr}].
In addition to recovering Bose statistics, Fermi statistics and
parastatistics (which is equivalent to internal color) they
found an additional, less familiar, possibility: ``infinite statistics''.
A collection of particles which obey infinite statistics can
be in any representation of the particle permutation group. These are
evidently the statistics obeyed by extremal black holes.

More recently
it was realized[\cite{grbg}] that that a Fock-like realization of
infinite statistics can be obtained from a $q$-deformation of
the commutation relations\footnote{*}{This should not be confused
with the $q$-deformed quantum theory of [\cite{igku}] in which
negative norm states appear.}:
$$
a_k a_l\d -qa_l\d a_k=\delta_{kl},
\eqno(qdef)
$$
where $k,~l$ may be viewed as labeling spatial momenta.
$q=\pm 1$ corresponds to bosons or fermions, while
other cases correspond to ``quons''.
States are built by acting on a vacuum which obeys
$$
a_k|0\rangle =0.
\eqno(vuu)
$$
It follows from \(qdef) and \(vuu) that
for $-1 < q <1$ one can form
$N!$ linearly independent states from $N$ oscillators
$a\d_{k_1}...a\d_{k_N}$ if the $k_i$ are all different.
(In the Bose or Fermi case there is only one independent state.)
This is easy to see for the special case $q=0$ for which
\(qdef) reduces to
$$
a_k  a_l\d=\delta_{kl}.
\eqno(zdef)
$$
It follows that the inner product of two $N$-particle states is
$$
\langle 0 |a_{k_N}...a_{k_1}a\d_{l_1}...a\d_{l_N}|0 \rangle
=\delta_{k_1 l_1}...\delta_{k_N l_N}.
\eqno(nrm)
$$
Thus two states obtained by acting with the $N$ oscillators in
different orders are orthogonal. Thus - as for extremal black holes -
the states may be in any representation of the permutation group.

It was noted in [\cite{grbg}] that the theory of quons
contains some mild forms of non-locality. For example the
expression for the Hamiltonian is both non-local and non-polynomial
in the field operators. Nevertheless cluster decomposition, CPT and
an analog of Wicks theorem were found to hold. It is not clear
if quons lead to physically unacceptable forms of
non-locality. The fact that quons naturally arise from a
sum over geometries suggests that it should be possible to
build a fully consistent theory. Perhaps the non-locality
discussed in [\cite{grbg}] is related to the fact that wormholes connect
spacelike separated points, and can be eliminated by
superselection rules arising from topological censorship.

In summary, we believe that understanding the quantum statistics of
extremal black holes is a key ingredient for unravelling quantum
black hole puzzles.  We have argued that, at least
in one description, extremal black holes
behave very differently from elementary particles. It is far from
obvious that this description is appropriate to the real world. Nevertheless
we feel it is a logical possibility - apparently consistent with quantum
mechanics - that our universe contains fundamental
entities which are neither fermions
nor bosons, but rather are quons.

\head{Acknowledgements}

I wish to thank S. Giddings, J. Harvey, A. Srivistava, S. Trivedi
and especially T.~Banks and J. Friedman for useful discussions. This work
was supported in part by DOE grant No. DEAC-03-8ER4050 and
NSF PHY89-04035.

\head{References}

\refis{igku} A. Yu. Ignatiev and V. A. Kuzmin, {\sl Yad. Phys. 46}
(1987) 786.

\refis{grbg} O. W. Greenberg, {\sl Phys. Rev. Lett. 64} (1990) 705;
{\sl Phys. Rev D43} (1991) 4111; J. Cuntz, {\sl Commun. Math. Phys. 57} (1977)
173. A recent summary can
be found in O. W. Greenberg, D. M. Greenberger and T. V. Greenbergest,
``(Para)fermions, (Para)bosons, Quons and Other Beasts in the Menagerie
of Particle Statistics'', talk presented at the
International Conference
on Fundamental Aspects of Quantum Theory (1993), hep-ph/9306225.

\refis{dhr} S. Doplicher, R. Haag and J. Roberts {\sl Commun. Math. Phys.
23} (1971) 199; {\sl Commun. Math. Phys.
35} (1974) 49.

\refis{gast}G.W. Gibbons, ``Quantized Flux Tubes in Einstein-Maxwell Theory
and  Noncompact Internal
Spaces,'' in {\sl Fields and geometry}, proceedings of
22nd Karpacz Winter School of Theoretical Physics,
Karpacz, Poland, Feb 17 - Mar 1, 1986, ed. A. Jadczyk (World
Scientific, 1986); D. Garfinkle and A. Strominger
{\sl Phys. Lett. B256} (1991) 146.

\refis{lyr} J. Preskill, P. Schwarz, A. Shapere, S. Trivedi and F. Wilczek,
\journal Mod. Phys. Lett., A6, 2353, 1991; C.G. Callan, S.B. Giddings,
J.A. Harvey and A. Strominger,
\pr D45, R1005, 1992; For recent reviews see J.A.
Harvey and A. Strominger,
``Quantum Aspects of Black Holes" preprint EFI-92-41, hep-th@xxx/9209055, to
appear in the proceedings of the 1992 TASI Summer School in Boulder, Colorado,
and S.B. Giddings, ``Toy Models for Black Hole Evaporation" preprint
UCSBTH-92-36, hep-th@xxx/9209113, to appear in the proceedings of the
International Workshop of Theoretical Physics, 6th Session, June 1992, Erice,
Italy.

\refis{bol} T. Banks, M. O'Loughlin and A. Strominger
hep-th/9211030, {\sl Phys.\ Rev.\ D \ 47 } (1993) 4476.

\refis{srdw}R. Sorkin and F. Dowker, to appear.

\refis{tc}J. Friedman, K. Schleich and D. Witt, ``Topological Censorship'',
ITP preprint NSF-ITP-93-80 (1993).

\refis{frb} D. Finkelstein and H. Rubinstein, {\sl J. Math Phys. 9}
(1968) 1762.

\refis{mty}M. S. Morris, K. S. Thorne
and U. Yurtsever,  {\sl Phys.\ Rev.\ Lett. \ 61 } (1988)
1446.

\refis{fsk}J. L. Friedman and R. D. Sorkin, {\sl Phys.\ Rev.\ Lett \ 44 }
(1980) 1100; {\sl Gen. Rel. And Grav.} (1982) 615. For a review
see J. L. Friedman ``Spacetime Topology and Quantum Gravity'' in
Proceedings of the Osgood Hill Conference on Conceptual Problems in
Quantum Gravity, eds. A. Ashketar and J. Stachel, Birkhauser (1991).

\refis{bal} An introduction to these ideas and their generalizations may be
found in the book ``Classical Topology and Quantum States'', World Scientific
(1991) by A. P. Balachandran, G. Marma, B. S. Skagerstam and A. Stern.

\refis{hwk} S. W. Hawking, {\sl Commun. Math. Phys. 43} (1975) 199,
{\sl Phys. Rev. D14} (1976) 2460, G. W. Gibbons,{\sl Commun. Math. Phys. 44}
(1975) 245.

\endreferences

\end